\documentclass[10pt,letterpaper,twocolumn]{article} 

\usepackage{ol2}
\usepackage{amsmath,amssymb,multirow}

\newcommand{\expt}[1]{\mathtt{E}\left[{#1}\right]}

\newcommand{\varn}[1]{\mathtt{var}\left[{#1}\right]}

\begin{document}

\twocolumn[ 

\title{Error analysis of moment-based modal wavefront sensing} 

\author{Hanshin Lee} 
\address{
McDonald Observatory, University of Texas at Austin, 2515 Speedway
C1402, Austin, TX, 78712\\ 
lee@astro.as.utexas.edu
}

\begin{abstract}
The shape of a focus-modulated point spread function (PSF) is used as a quick visual assessment tool of aberration modes in the PSF. Further analysis in terms of shape moments can permit quantifying the modal coefficients with an accuracy comparable to that of typical wavefront sensors. In this letter, the error of the moment-based wavefront sensing is analytically described in terms of the pixelation and photon/readout noise. All components highly depend on the (unknown) PSF shape, but can be estimated from the measured PSF sampled at a reasonable spatial resolution and photon count. Numerical simulations verified that the models consistently predicted the behavior of the modal estimation error of the moment-based wavefront sensing. \\ 
\end{abstract}


] 
\noindent 
\textbf{Introduction:}
The shape of a focus-modulated PSF changes in certain patterns that can be attributed to different wavefront modes. Such shape variation is an excellent tool for visually assessing the image quality of optical systems~\cite{Suiter}. Among many PSF shape parameters, the width, ellipticity, and diffraction rings are used to assess focus, astigmatism, and spherical aberrations. However, the PSF can be complex in shape and sampled over a pixel grid much coarser than the diffraction scale. This makes it challenging to readily measure the aforesaid shape parameters. Alternatively, one can use phase diversity~\cite{Fineup} or curvature sensing~\cite{RoddierCS}. 
Each technique has its own intrinsic strength in particular applications over the other~\cite{Hardy} and effort has been made to broaden the application of these techniques~\cite{Gonsalves,Keller,Dolne,LIFT}. 
 
Meanwhile, some explored a different approach to focal-plane wavefront sensing, where the dependence of PSF shape on wavefront modes is utilized in the form of a linear relation between the PSF shape moments and wavefront modal coefficients. Teague first recognized this moment-aberration relation in the context of developing the intensity transport equation aspect of phase retrieval problem~\cite{Teague}. Recently, an independent study also recognized the relation in terms of in-situ fine alignment of astronomical spectrographs~\cite{Lee1}. 
Both have shown that the linear relation between aberration modes and moments permits non-iterative determination of modal coefficients at various orders. The latter also reported the moment-based sensing accuracy can be comparable to that of typical wavefront sensors. Also noted by Teague and shown elsewhere~\cite{Lee2} is the point source formulation being applicable to extended objects. A field application has been reported~\cite{Lee3}. In this letter, the error of the moment-based wavefront sensing is analyzed in terms of the pixelation error, photon noise, and read-out noise.

\noindent \textbf{Theory:}
To begin, the basic theory of the moment-based wavefront sensing~\cite{Lee1} is revisited. Let $\theta(s,t)$ be the wavefront aberration from a point object at a fixed field and defined over a circular pupil $\Omega$ whose coordinate is ($s,t$). 
Let $\theta$ be expressed as a weighted sum of $M$ Zernike modes (Eq.~\ref{zernike}) and $W_4$ (i.e. the 4th modal coefficient) be modulated as in focusing through a detector.
\begin{align}
	\theta=W_1Z_1 + W_2Z_2 + W_3Z_3 + \cdots + W_{M}Z_{M}.
	\label{zernike}
\end{align}
Noll's Zernike index scheme in~\cite{Noll} used throughout. 
$\theta$ results in ray aberrations around the ideal image as,
\begin{align}
	X = -f\,{\partial_{s}\theta} = -f \theta_s\;\;\mathrm{and}\;\;
	Y = -f\,{\partial_{t}\theta} = -f \theta_t,
	\label{rayabr}
\end{align}
\noindent where {$f$} is the focal length and {$\partial_a\theta = \theta_a$ is the partial derivative of $\theta$ with respect to $a$. Note that $\theta_s$ and $\theta_t$ are linear functions of $W_4$ and so are $X$ and $Y$.}

The total signal within a section of the focal surface can be given by summing the irradiance of all \textit{geometric} rays landing there. The collection of the signals of these sections forms the PSF whose shape can be characterized by the $q$-th shape moment  as,
\begin{align}
	\mu^{g}_{nm} = \Big\{\iint_{\Omega}I\;X^n\;Y^m \mathrm{d}s\mathrm{d}t \Big\}/
	\Big\{\iint_\Omega I\,\mathrm{d}s\mathrm{d}t\Big\},
	\label{moment}
\end{align}
\noindent where $I$ is the irradiance over $\Omega$ and {\textit{q=n}+\textit{m} with \textit{n,m}$\geq$0}. {$\mu^g_{nm}$ is the \textit{geometric} version of the PSF shape moment in the pupil plane}. {In the focal plane,}
\begin{align}
	\mu^d_{nm} = \frac{\iint_{\Lambda}p\;x^n y^m \mathrm{d}x\mathrm{d}y}{\iint_\Lambda p \;\mathrm{d}x\mathrm{d}y} \approx
	\mathtt{M}_{nm}= \frac{\sum\nolimits_{k=1}^{N_{px}}p_k x_k^n y_k^m}{ \sum\nolimits_{k=1}^{N_{px}} p_k},
	\label{moment_fp}
\end{align}
\noindent where $p$ is the signal within the PSF region, $\Lambda$. \texttt{M}$_{nm}$ emphasizes that $\mu^d_{nm}$ can only be approximated by discrete signal $p_k$ in the $k$-th pixel at ($x_k$,$y_k$).
Despite that $\mu^d_{nm}$ differs from $\mu^g_{nm}$ due to $p$ being partly given by \textit{diffraction}, Eq.~\ref{moment} and~\ref{moment_fp} are assumed equivalent for the purpose of estimating $W_k$. Thus $\mu_{nm} \equiv \mu^g_{nm} \equiv \mu^d_{nm}$ hereafter.

Since $X$ and $Y$ are linear functions of $W_4$, $\mu_{nm}$ becomes a polynomial of order $q$ in $W_4$. The highest order term is $W_4$ to the power of $q$ (i.e. $W_4^{q})$ whose polynomial coefficient must be a constant. The next highest order term is $W_4^{q-1}$. Its coefficient must be a weighted linear sum of some $W_k$ that are sensitive to the variation in this image moment, which is the key to establishing the linear relation between $W_k$ and $\mu_{nm}$. This \textit{q-1}th coefficient is equivalent to $(\partial/\partial W_4)^{q-1}\mu_{nm}$ (to be called $d\mu_{nm}$).

To measure $d\mu_{nm}$, $N$ focus-modulated focal plane images are recorded and the moments of order up to $q$ are computed from the images. The $q$-1th coefficient of an order $q$ polynomial fit to the modulated moments is used as $d\mu_{nm}$. For the unique polynomial fit, $N \geq q+1$. Note there are $q$+1 $d\mu_{nm}$ at each order and, up to order $q$, there are $L=q(q+3)/2$ $d\mu_{nm}$. One can formulate,
\begin{align}
	\vec{u}=\mathbf{M}\vec{W}.
	\label{moment_solve}
\end{align}
\noindent \textbf{M} is a matrix given by the integral in Eq.~\ref{moment} for $W_4^{q-1}$ terms. $\vec{u}$ and $\vec{W}$ are vectors of $d\mu_{nm}$ and $W_k$, respectively. Solving Eq.~\ref{moment_solve} results in the first $L$ $W_k$ (no $W_1$).

\begin{figure}[htbp]
	\centering
	\includegraphics[width=83mm]{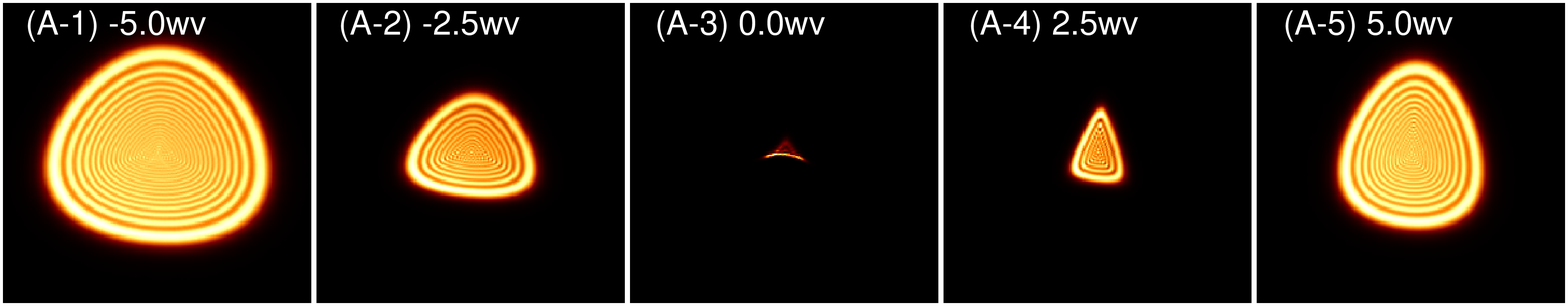}
	\includegraphics[trim=0cm 0cm 0cm 0cm,clip=true,width=83mm]{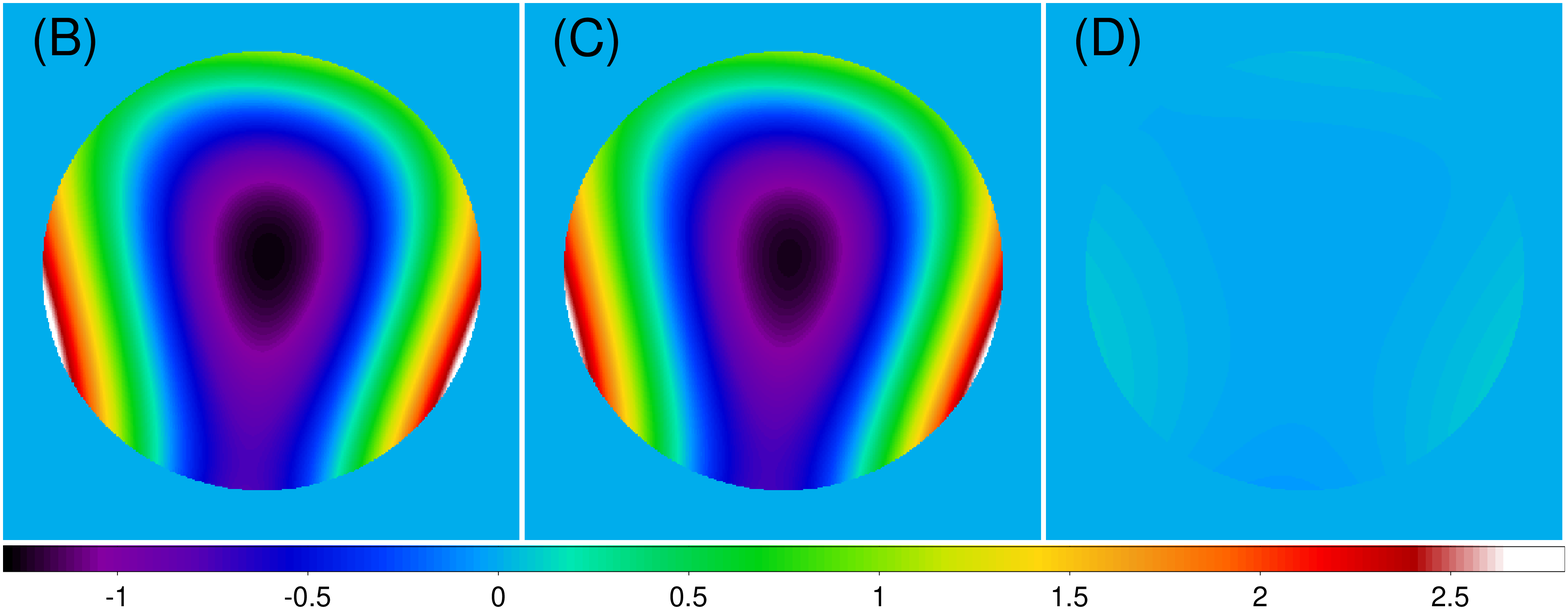}
	\caption{\footnotesize{5 modulated PSF images (A) and phase maps (B-D). The pixel size in (A) is 5$\mu m$. The color bar unit is wv.}}
	\label{Figure0}
\end{figure}

To validate the assumption of $\mu_{nm}^g \equiv \mu_{nm}^d$, PSFs without photon and read noise are synthesized for 5 focus-modulations using the Fourier method~\cite{Fineup} (Fig.~\ref{Figure0}-A). The input phase in (B), given by the modes from $W_2$ to $W_9$ and of 0.92wv in root-mean-square (rms), is used for the PSF synthesis. By applying the aforesaid procedures, the phase estimate in (C) is obtained and its difference from (B) is shown in (D). The rms of this residual is 0.025wv. Note beam speed of f/8 and wv=632.8nm.

\noindent \textbf{Error models:}
Since the PSF is a photon distribution recorded on a pixelated detector, its shape moment is subject to the pixelation error. Given the total PSF photons ($N_{ph}$) and letting $p(x,y)$ be the true photon distribution, the photon count ($p_k$) is the integral of $p$ over at the k-th pixel ($\Delta_k$). By letting $\delta x_k=x-x_k$ and $\delta y_k=y-y_k$ and by expanding the moment kernel $\phi_{nm}(x,y)=x^n y^m$ to a series around $(x_k,y_k)$, $\mu_{nm}$ in Eq.~\ref{moment_fp} can be written as,
\begin{align}
	\mu_{nm} = \frac{1}{N_{ph}} \sum_{k=1}^{N_{px}}
	\iint_{\Delta_k}& p\, \phi_{nm}+
	p\,(\partial_x\phi_{nm})\,\delta x_k+ \nonumber \\
	& p\,(\partial_y\phi_{nm})\,\delta y_k + {O(\epsilon)}\;
	\mathrm{d}x\mathrm{d}y.
	\label{Ga3}
\end{align}
\noindent 
Here $\phi_{nm}$, $\partial_x\phi_{nm}$, and $\partial_y\phi_{nm}$ are given at (\textit{x$_k$,y$_k$}) and $O(\epsilon)$ includes all higher order terms. The first term is $\mathtt{M}_{nm}$. The rest is the pixelation error ($\delta_{nm}$) of $\mathtt{M}_{nm}$. Expanding $p$ in $\delta_{nm}$ into a series within $\Delta_k$ leads to, 
\begin{align}
	\delta_{nm} \approx -\frac{\Delta^2}{12} \sum_{k=1}^{N_{px}}
	\frac{
	\partial_x\phi_{nm}\partial_x p+\partial_y\phi_{nm}\partial_y p}{N_{ph}/\Delta^2}, 
	\label{Ga5}
\end{align}
\noindent where $\Delta$ is the pixel size. The exact value of $\delta_{nm}$ depends on the (unknown) PSF shape. But, one can gain the following insights. The sum can be viewed as a 2-dimensional integral of $\nabla\phi_{nm}\cdot\nabla p$. Like the one in Eq.~\ref{moment_fp}, this integral quantifies the response of $\nabla p$ to $\nabla \phi_{nm}$. Intuitively, if $p$ has a strong response to a certain $\phi_{nm}$ (thus large $\mu_{nm}$), its gradient should respond to the gradient of the same kernel in a similar way if not exactly the same, implying a strong correlation between $\delta_{nm}$ and $\mu_{nm}$ (e.g. $|\delta_{nm}| \approx 0$ for $|\mu_{nm}|=0$). Another aspect is $\delta_{nm} \propto \Delta^2$, confirming the known fact that a higher spatial resolution (or $N_{px}$) leads to a smaller $\delta_{nm}$. Finally, the pixelation error is due to a lack of approximation to the true PSF and thus systematic. In Eq.~\ref{Ga5}, one may use the numerical gradient of $p$ and $\phi_{nm}$ to estimate $\delta_{nm}$, but, since $\phi_{nm}$ is known, a better estimate may be obtained using all derivatives of $\phi_{nm}$, as,
\begin{align}
	\delta_{nm} \approx \frac{1}{N_{ph}}&\sum_{k=1}^{N_{px}}\bigg[\sum_{l=1}^q\bigg\{\sum_{i=0}^{l}
	\frac{1}{i!j!}{\partial^i_x\partial^j_y\phi_{nm}}\bigg(
	\iint_{\Delta_k}p_0 \mathtt{x}^i \mathtt{y}^j + \nonumber \\
	& {(\partial_x p)}\, \mathtt{x}^{i+1} \mathtt{y}^j +
	{(\partial_y p)}\, \mathtt{x}^i \mathtt{y}^{j+1}\mathrm{d} \mathtt{x}\mathrm{d} \mathtt{y} \bigg)\bigg\}\bigg].
	\label{Enm}
\end{align}
\noindent Here $j=l-i$, $\mathtt{x}=x_k+x$ and $\mathtt{y}=y_k+y$, and $p_0$ and all derivatives are evaluated at ($x_k,y_k$). 

Besides the pixelation error, the intrinsic uncertainty of photon detection also imposes Poisson random error $P(\cdot)$ and the imaging detector adds zero mean Gaussian random read noise $G(\cdot)$ to each pixel value, leading to random error in $\mathtt{M}_{nm}$. Let the value of the \textit{k}th pixel be $s_k = P(p_k) + G(\sigma)$, where $p_k$ is the mean photon count and $\sigma$ is the read noise in rms. The expectation of the estimate $\hat{\mathtt{M}}_{nm}$ is given as,
\begin{align}
	\expt{\hat{\mathtt{M}}_{nm}} = &  \expt{
	\frac{\sum_k s_k \phi_{nm,k}}{\sum_k s_k} } = \expt{\frac{A}{B}}
	= \expt{A} \expt{\frac{1}{B}} \nonumber \\
	\approx & \mathtt{M}_{nm} ( 1 + {\mathtt{SNR}^{-2}} ),
	\label{Mean}
\end{align}
\noindent where $\mathtt{SNR}^2={(\sum_k p_k)^2}/{\sum_k (p_k + \sigma^2)}$. $A$ and $1/B$ are assumed independent. The series expansion of $1/B$ up to order 2 is used to compute $\expt{1/B}$.
The variance of $\hat{\mathtt{M}}_{nm}$ is expressed as $\varn{\hat{\mathtt{M}}_{nm}} = 
	\expt{A^2} \expt{1/B^2} - \expt{\hat{\mathtt{M}}_{nm}}^2$,
where no correlation between $A^2$ and $1/B^2$ is assumed. Expanding $1/B^2$ as done for $1/B$ in Eq.~\ref{Mean} leads to
\begin{equation}
	\varn{\hat{\mathtt{M}}_{nm}} \approx
	\frac{{\mathtt{M}}_{nm}^2}{\mathtt{SNR}^2} +
		\frac{\sum_k p_k\phi_{nm,k}^2}{(\sum_k p_k)^2}+
		\frac{\sigma^2\sum_k\phi_{nm,k}^2}{(\sum_k p_k)^2}\!.
	\label{varn1}
\end{equation}

\noindent The first term grows with the moment and is scaled by $1/\mathtt{SNR}^2$ (or $1/N_{ph}$). Since $\phi_{nm,k}^2 = x_k^{2n} y_k^{2m}$, the second term is essentially ${\mathtt{M}}_{2n2m}$ scaled by $1/N_{ph}$. For instance, the variance of ${\mathtt{M}}_{10}$ of a PSF with a finite width is given by ${\mathtt{M}}_{20}$, which is non-zero even if the PSF is perfectly centered. These two essentially depend on the intrinsic shape of the (unknown) PSF. The last term is due to $\sum_k\phi_{nm,k}^2 \propto N_{px}H^{2(n+m)}$, where $H$ is a characteristic size of the PSF region (e.g. max. radius), scaled by $\sigma^2/N^2_{ph}$. For a PSF with fixed $N_{px}$ (or resolution) and $\sigma$, this term follows a $1/N^2_{ph}$ trend and can dominate the other two terms especially at low \texttt{SNR}. For a bright target at a high resolution, these terms could be estimated by substituting $\mathtt{\hat{M}}_{nm}$ and $s_k$ for $\mathtt{{M}}_{nm}$ and $p_k$, respectively.

The error in Eq.~\ref{varn1} then propagates to the $q\times$1 coefficient vector $\vec{c}$ of an order $q$ polynomial fit to $\{\hat{\mathtt{M}}_{nm}\}$ measured at focus modulations $\{F_i, \,i=1,2,\cdots,K\}$ (Eq.~\ref{fit}).
\begin{align}	
	\vec{c}=(\mathbf{A}^T\mathbf{A})^{-1}(\mathbf{A}^T\vec{b})\;\mathrm{with}\;
	A_{ij}=\frac{F_i^j}{\mathtt{e}_i}\;,\;
	b_{i}=\frac{\mathtt{\hat{M}}^{(F_i)}_{nm}}{\mathtt{e}_i}.
	\label{fit}
\end{align}

\noindent where $\vec{b}$ is the $K\times 1$ vector of \{$\hat{\mathtt{M}}_{nm}$\}, $\mathbf{A}$ is a $K\times q$+1 fit matrix, and $\mathtt{e}_i$ is the square root of $\varn{\mathtt{\hat{M}}_{nm}}$ at $F_i$. $c_{q}$ becomes the estimate of $d\mu_{nm}$. Letting $\mathbf{D}=(\mathbf{A}^T\mathbf{A})^{-1}$, $D_{qq}$ becomes the error variance estimate of $d\mu_{nm}$~\cite{NRC} and propagates to $\vec{W}$ as,
\begin{align}	
	\vec{W}=(\mathbf{B}^T\mathbf{B})^{-1}(\mathbf{B}^T\,\vec{d})\;\mathrm{with}\;
	B_{ij}=\frac{M_{ij}}{\mathtt{v}_i}\;,
	d_{i}=\frac{u_{i}}{\mathtt{v}_i},
	\label{fit2}
\end{align}
\noindent where $\mathtt{v}_i$ is the square root of the error variance of $u_i$. Finally, by Letting $\mathbf{J}=(\mathbf{B}^T\mathbf{B})^{-1}$, $J_{ii}$ becomes the error variance estimate of $W_{i}$. 

\begin{figure}[tp]
	\centering
	\includegraphics[width=83mm]{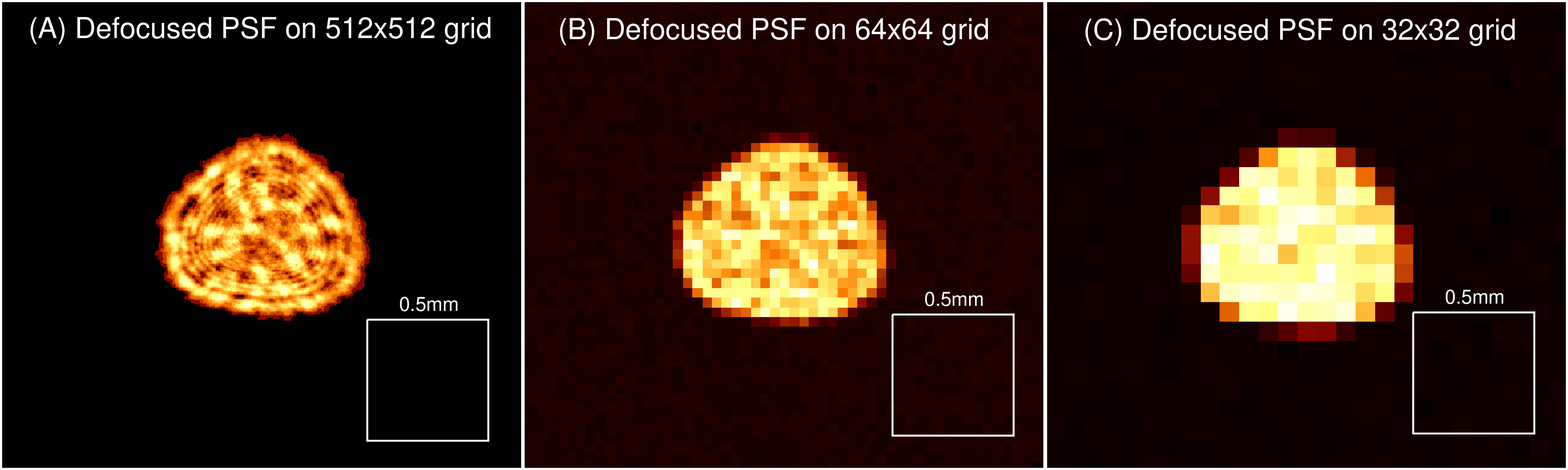}
	\includegraphics[width=83mm]{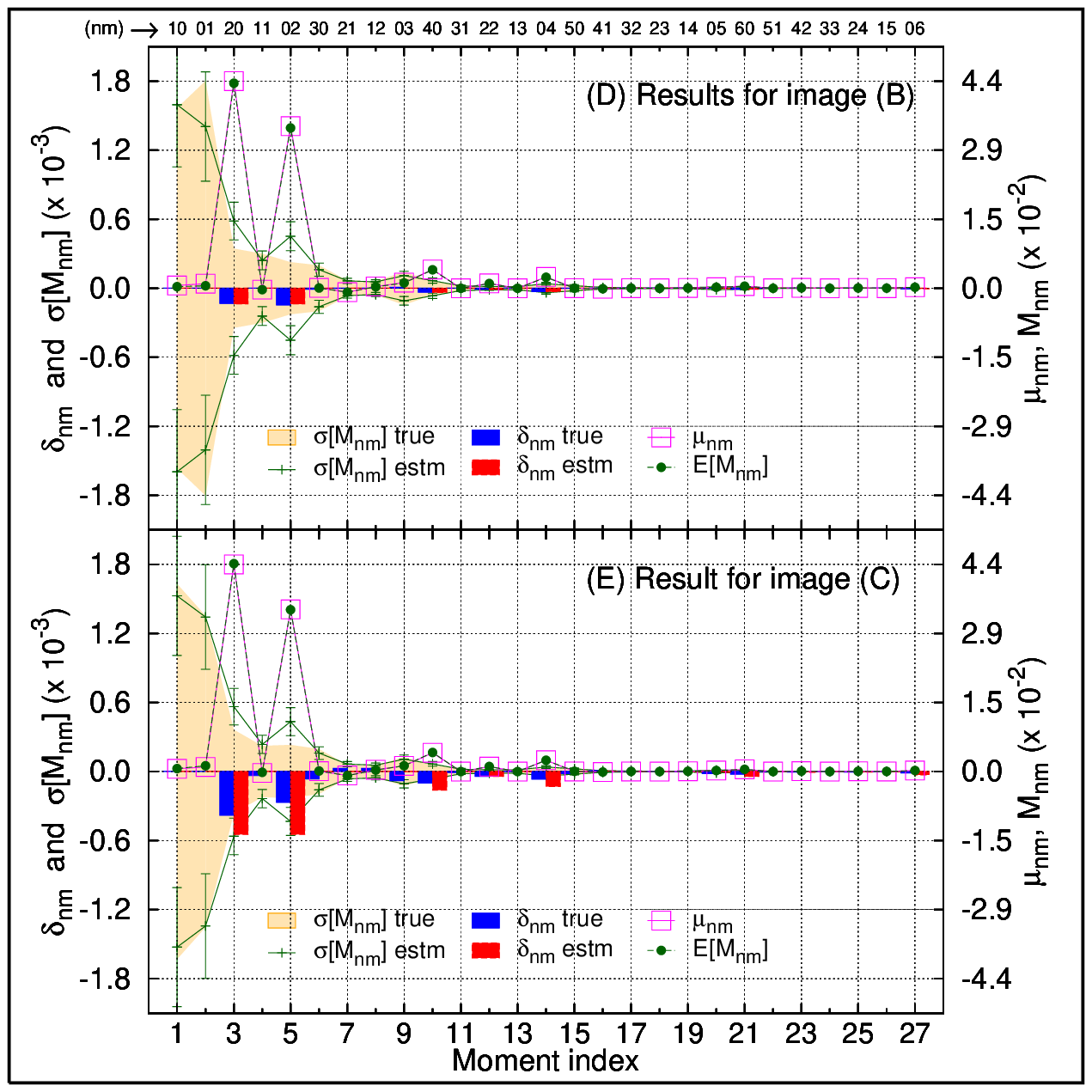} 	
	\caption{\footnotesize{A defocused PSF at different resolutions (A-C). Error analysis results for (B) and (C) (D,E). 5$\mu$m pixel in (A).}}
	\label{Figure1}
\end{figure}

\noindent\textbf{Simulations:} The error models are examined for the defocused PSF images shown at the top of Fig.~\ref{Figure1}. The first 27 moments of (A) are set as the true moments. For each under-sampled image in (B) and (C), 50 random frames were created according to $P(p_k)$ and $G$($\sigma$=3e$^-$) with $N_{ph} = 20,000$e$^-$, and the followings in Table~\ref{table1} have been computed and plotted in (D) and (E) for (B) and (C), respectively. $s_k (< 5\sigma)$ was ignored in the analysis.
\vspace{-0.0cm}

\begin{table}[htbp]
\small
\centering
\caption{ Quantities computed and shown in Fig.~\ref{Figure1} (D,E).}
\begin{tabular}{c|c}
\hline\hline
	Computed quantity & Plot symbol \\ \hline
	$\mu_{nm}$ = Moments of (A) & $\mu_{nm}$ \\ \hline
	$\expt{\mathtt{M}_{nm}}$ = Mean across 50 $\{\hat{\mathtt{M}}_{nm}\}$ & $\expt{\mathtt{M}_{nm}}$ \\ \hline
	$\delta_{nm}$ = $\mu_{nm} - \expt{\mathtt{M}_{nm}}$ & $\delta_{nm}$ \texttt{true}  \\ \hline
	Estimate of $\delta_{nm}$ using Eq.~\ref{Enm} & $\delta_{nm}$ \texttt{estm} \\ \hline
	$\sqrt{\varn{\mathtt{\hat{M}}_{nm}}}$ = 1$\sigma$ across 50 $\{\hat{\mathtt{M}}_{nm}\}$ & $\sigma[\mathtt{M}_{nm}]$ \texttt{true} \\ \hline
	Estimate of $\sqrt{\varn{\mathtt{\hat{M}}_{nm}}}$ using Eq.~\ref{varn1} & $\sigma[\mathtt{M}_{nm}]$ \texttt{estm}  \\ \hline\hline
\end{tabular}
	\label{table1}
\end{table}

As discussed in Eq.~\ref{Ga5}, larger $|\mu_{nm}|$ tends to accompany larger $|\delta_{nm}|$ at least within the same order. The moments with negligible power show near zero $\delta_{nm}$. Equation~\ref{Enm} appears approximating $\delta_{nm}$ to a reasonable extent for the 64$\times$64 grid, but the approximation becomes poorer for the lower resolution. $|\delta_{nm}|$ also appears to be roughly quadrupled from (B) to (C) as discussed in Eq.~\ref{Ga5}. Meanwhile, $\mathsf{\sigma[{{M}_{nm}}]}$ \textsf{estm} closely follows the true variation ($\mathsf{\sigma[{{M}_{nm}}]}$ \textsf{true}) in both resolutions. Note the large variations in $\mu_{10}$ and $\mu_{01}$ (index 1 and 2) due to large $\mu_{20}$ and $\mu_{02}$ (index 3 and 5).
The error bar shows the 1$\sigma$ variation in estimating $\mathsf{\sigma[{{M}_{nm}}]}$ across the 50 realizations.

\begin{figure*}[t]
	\centering 
 	\includegraphics[trim=0cm 0.8cm 0cm 0cm,clip=true,width=35.5mm]{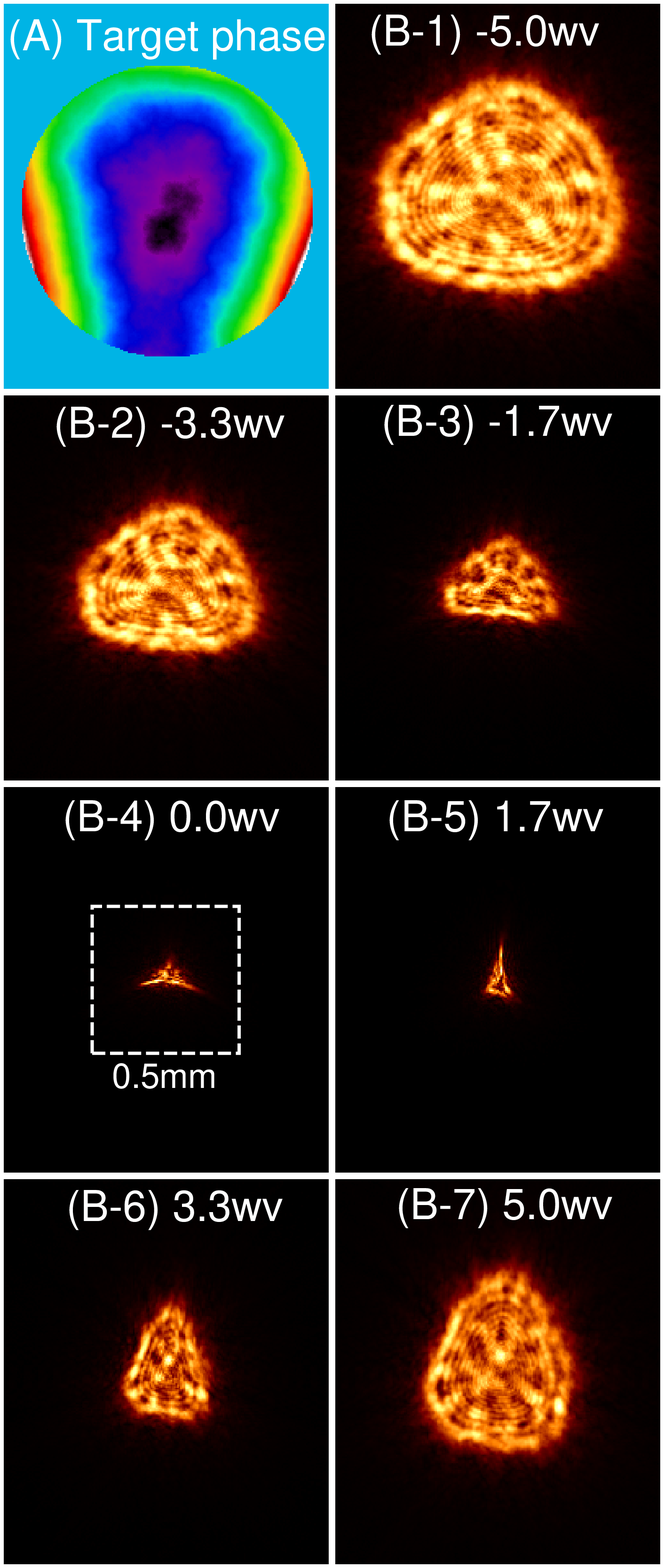}
	\includegraphics[trim=0cm 0.45cm 0cm 0.0cm, clip=true,width=129mm]{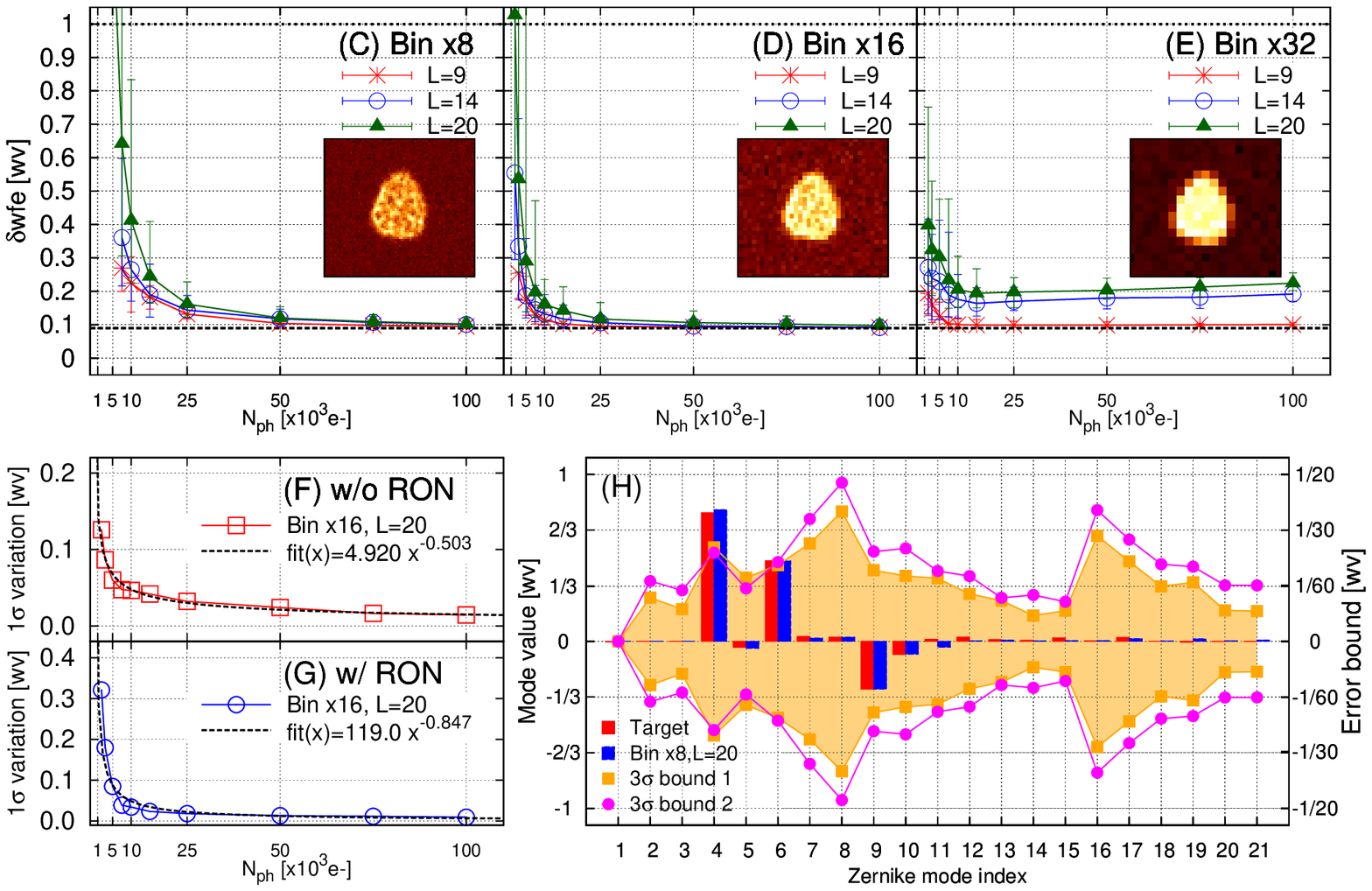}
	\caption{\footnotesize{(A) Target phase and (B) Modulated PSF images used in the moment-based modal sensing error analysis (C-H).}}
	\label{Figure2}
\end{figure*}

In Fig.~\ref{Figure2}, the modal sensing error is examined. The target phase is shown in (A). The phase is given by a Kolmogorov phase screen of D/$r_0$=10 (D is the pupil diameter and $r_0$ is Fried parameter~\cite{Hardy}). The screen was generated by using the method in~\cite{Kol} and most of its power is contained in the first 10 Zernike terms. The phase is used to synthesize 7 focus-modulated PSFs (between $\pm5$wv) using the same method as in Fig.~\ref{Figure0} for f/8 beam. These PSFs are shown in (B). 

The phase is estimated by the moment method and the rms of the difference ($\mathsf{\delta wfe}$) between the estimate and the target is shown in Plot (C-E) for different resolutions (Bin), the number of photons per modulation ($N_{ph}$), and the number of terms estimated ($L$). Each point represents the mean $\delta \mathsf{wfe}$ of 101 random cases and the error bar brackets the variation. The read noise is fixed to 3e$^-$ and 5$\sigma$ cut-off was applied. The pixel size is 20$\mu m$ at Bin$\times 8$. In each plot, the inset shows the PSF at 5wv modulation. The dotted horizontal line at 0.964wv indicates the rms aberration of the target and the dashed line at 0.092wv shows the lower limit set by the un-sensed high frequency phase that is mostly in $Z_{22}$ and higher.

The curves asymptote ${N^{-1}_{ph}}$ (or \texttt{SNR}$^{-2}$) as discussed in Eq.~\ref{Mean}. 
A larger bin increases \texttt{SNR}, reducing $\delta \mathsf{wfe}$ quickly to the lower limit even at low ${N_{ph}}$, especially for $L$=9. The penalty is an elevated lower limit at high ${N_{ph}}$ due to larger systematic $\delta_{nm}$ affecting the higher order moments and the estimates for $L$=14 and 20. Such effect can be suppressed by increasing resolution, but at the cost of lower \texttt{SNR}. Note in (D) that the mean $\delta$\textsf{wfe} across the realizations is 0.097wv at $N_{ph}$=100,000e$^-$. Considering the un-sensed phase of 0.092wv, the mean estimation residual amounts to 0.031wv. 

Plot (F-G) show the 1$\sigma$ variation in estimating the target phase without and with the read noise, respectively. As discussed in Eq.~\ref{varn1}, the photon-noise limited variation in (F) follows $N^{-0.5}_{ph}$. With the read noise, the variation in (G) asymptotes $N^{-1}_{ph}$.

Plot (H) shows the first 21 Zernike coefficients of the target (red bars) and estimated (blue bars) phase for Bin$\times8$ and $N_{ph}$=100,000e$^-$ in (A). The estimate correctly captured the major modal contents of the target phase. Also shown is the 3$\sigma$ error bounds, one derived from the 101 random sets (``$3\sigma$ bound 1'') and the other computed from the models in Eq.~\ref{fit} and~\ref{fit2} (``$3\sigma$ bound 2''). The models over-estimate the actual variation, but appear to have correctly captured the overall bounds.

\noindent \textbf{Conclusion:} The error analysis of the moment-based modal wavefront sensing is presented along with the basic theory behind the method. The systematic pixelation and the random error due to photon and read noise are modeled. The model prediction appears consistent with the numerical simulations, demonstrating that one can use the models to find the optimal parameter space where the moment method can be effective. Also it is shown that the estimation residual can approach 0.03wv, when $N_{ph}$ and resolution are reasonably high (e.g. as in lab settings) and target phase is mostly of low-order. 

In principle, sensing higher-order mode should be possible, but the increased noise sensitivity of higher-order moments would become a challenge. When large higher-order modes exist, the aliasing effect into lower-order becomes important and needs a further investigation. 

Given routine extra-focal imaging and the moment computation being an extension of centroid calculation, the moment method can be implemented with minimal hardware/software effort for everyday image quality assessment tasks. Its low-order estimate could also be a rapid quality initial guess to accelerate a follow-up fine phase retrieval by e.g. phase diversity on the same PSFs.

\pagebreak

\end{document}